\documentclass[10pt,letterpaper]{article}
\usepackage[top=0.85in,left=2.75in,footskip=0.75in]{geometry}

\usepackage{amsmath,amssymb}

\usepackage{changepage}

\usepackage{textcomp,marvosym}

\usepackage{cite}

\usepackage{nameref,hyperref}
\usepackage{booktabs}
\usepackage{threeparttable}
\usepackage{adjustbox}
\usepackage{url}

\usepackage[right]{lineno}

\usepackage[nopatch=eqnum]{microtype}
\DisableLigatures[f]{encoding = *, family = * }

\usepackage[table]{xcolor}

\usepackage{array}
\usepackage{textcomp,marvosym}
\usepackage[utf8]{inputenc}
\usepackage[T1]{fontenc}
\usepackage{float}
\usepackage{algorithm}
\usepackage{algpseudocode}
\usepackage{float}
\usepackage{tabularx}
\usepackage{booktabs}
\usepackage{placeins}
\usepackage{cite}
\newcolumntype{+}{!{\vrule width 2pt}}

\newlength\savedwidth

\raggedright
\usepackage[aboveskip=1pt,labelfont=bf,labelsep=period,justification=raggedright,singlelinecheck=off]{caption}

\makeatletter
\renewcommand{\@biblabel}[1]{\quad#1.}
\makeatother

\usepackage{lastpage,fancyhdr,graphicx}
\usepackage{epstopdf}
\fancyheadoffset[L]{2.25in}
\fancyfootoffset[L]{2.25in}
\begin{document}
\vspace*{0.2in}

\begin{flushleft}
{\Large
\textbf\newline{Predicting Postoperative Stroke in Elderly SICU Patients: An Interpretable Machine Learning Model Using MIMIC Data} 
}


Tinghuan Li\textsuperscript{1},
Shuheng Chen\textsuperscript{1},
Junyi Fan\textsuperscript{1},
Elham Pishgar\textsuperscript{2},
Kamiar Alaei\textsuperscript{3},
Greg Placencia\textsuperscript{4},
Maryam Pishgar\textsuperscript{1*}
\textsuperscript{\textpilcrow}
\\
\bigskip
\textbf{1} Department of Industrial and Systems Engineering, University of Southern California, 3715 McClintock Ave GER 240, Los Angeles, 90087, California, United States
\\
\textbf{2} Colorectal Research Center, Iran University of Medical Sciences, Tehran Hemat Highway next to Milad Tower, Tehran, 14535, Iran
\\
\textbf{3} Department of Health Science, California State University, Long Beach (CSULB), 1250 Bellflower Blvd, Long Beach, 90840, California, United States
\\
\textbf{4} Department of Industrial and Manufacturing Engineering, California State Polytechnic University, Pomona, 3801 W Temple Ave, Pomona, 91768, California, United States
\\
\bigskip

\textpilcrow Membership list can be found in the Acknowledgments section.

* pishgar@usc.edu

\end{flushleft}
\section*{Abstract}
Postoperative stroke remains a critical complication in elderly surgical intensive care unit (SICU) patients, contributing to prolonged hospitalization, elevated healthcare costs, and increased mortality. Accurate early risk stratification is essential to enable timely intervention and improve clinical outcomes. We constructed a combined cohort of 19,085 elderly SICU admissions from the MIMIC-III and MIMIC-IV databases and developed an interpretable machine learning (ML) framework to predict in-hospital stroke using clinical data from the first 24 hours of Intensive Care Unit (ICU) stay. The preprocessing pipeline included removal of high-missingness features, iterative Singular Value Decomposition (SVD) imputation, z-score normalization, one-hot encoding, and class imbalance correction via the Adaptive Synthetic Sampling (ADASYN) algorithm. A two-stage feature selection process—combining Recursive Feature Elimination with Cross-Validation (RFECV) and SHapley Additive exPlanations (SHAP)—reduced the initial 80 variables to 20 clinically informative predictors. Among eight ML models evaluated, CatBoost achieved the best performance with an AUROC of 0.8868 (95\% CI: 0.8802--0.8937). SHAP analysis and ablation studies identified prior cerebrovascular disease, serum creatinine, and systolic blood pressure as the most influential risk factors. Our results highlight the potential of interpretable ML approaches to support early detection of postoperative stroke and inform decision-making in perioperative critical care.


\section*{Introduction}
Stroke is a clinical syndrome characterized by an acute disturbance in localized brain function, typically resulting from vascular events such as arterial occlusion or hemorrhage. This disruption in cerebral blood flow can cause prolonged neurological impairment, commonly exceeding 24 hours, and induces neuronal damage due to insufficient oxygen delivery~\cite{kuriakose2020pathophysiology,kelly2010influence}. Its underlying pathophysiology involves either acute ischemia or hemorrhage, followed by secondary injury cascades—such as excitotoxicity, oxidative stress, and neuroinflammation—that lead to progressive neurological and functional decline~\cite{kuriakose2020pathophysiology}. Ischemic strokes, caused by vascular obstruction, account for nearly 87\% of all strokes, while the remainder result from hemorrhagic events involving intracranial bleeding~\cite{donnan2008stroke}. Globally, stroke represents a major contributor to long-term disability and premature death, affecting over 13 million people annually and leading to approximately 5.5 million fatalities~\cite{whoemro_stroke}. Beyond its clinical impact, stroke presents a substantial economic challenge. In the United States, total costs reached \$103.5 billion, with 66\% attributed to indirect expenses such as productivity loss and premature mortality \cite{girotra2020costs}. In Europe, 2017 estimates reported \texteuro27 billion in healthcare costs, \texteuro12 billion from lost productivity, and \texteuro1.3 billion for informal care \cite{luengo2020economic}.

Although stroke incidence among younger adults, defined here as those between 20 and 54 years old, has shown a notable rise from approximately 13\% to nearly 19\% over the period spanning 1990 to 2016~\cite{GBD2019stroke}, the majority of stroke cases and the greatest burden of stroke-related disability and mortality remain concentrated in the elderly population. In fact, for individuals aged over 55 years, the likelihood of experiencing a stroke roughly doubles with each successive decade, driven largely by age-related vascular decline, common comorbidities such as hypertension and atrial fibrillation, and diminished physiological resilience~\cite{kelly2010influence,boehme2017stroke}. Notably, over 75\% of all strokes occur in individuals aged 65 and older~\cite{yousufuddin2019aging}. Additionally, the reported stroke prevalence in this population has been adjusted to account for differences in age and sex distribution, resulting in values ranging from 4.6\% to 7.3\% based on a meta-analysis of nine population-based studies~\cite{feigin2003stroke}. In Europe alone, stroke-related disability affects 2.7 million older adults, with an estimated 536{,}000 new cases and 1.24 million deaths annually—including 508{,}000 within the European Union~\cite{di2009human}. Despite advances in perioperative management, stroke remains a severe postoperative complication among elderly patients undergoing major surgery~\cite{mantz2010outcomes,biteker2014impact,dong2017risk}. Existing clinical strategies—such as antiplatelet therapy, blood pressure control, and rehabilitation—are predominantly reactive, offering limited value in predicting or preventing stroke onset~\cite{kamel2018prevention,catanese2017stroke,towfighi2011stroke}. With the global population aged 65 and older projected to surpass that of adolescents and young adults aged 15--24 by 2050~\cite{Grinin2023GlobalAging}, the burden of postoperative stroke is expected to rise substantially, placing growing pressure on critical care and rehabilitation systems. These trends underscore the urgent need for proactive, data-driven tools to identify high-risk individuals prior to stroke occurrence and support timely, targeted interventions.

Whitlock et al.~(2014)~\cite{whitlock2014predictors} conducted a population-based cohort study to evaluate early and late stroke risk among patients undergoing cardiac surgery. Traditional statistical approaches, including Cox proportional hazards models and the CHADS$_2$ risk score, were used to identify predictors of postoperative stroke. The study found that advanced age ($\geq$ 65 years), prior stroke or transient ischemic attack, peripheral vascular disease, and complex surgical procedures were significantly associated with increased stroke risk. The CHADS$_2$ score effectively stratified long-term stroke risk among patients with preexisting or new-onset atrial fibrillation.

Peguero et al.~(2015)~\cite{peguero2015cha2ds2vasc} retrospectively analyzed a cohort of 3,492 cardiac surgery patients to determine clinical factors linked to the development of postoperative ischemic stroke. Univariate analyses (e.g., t-tests and chi-square tests) were first applied to identify significant associations between baseline characteristics and stroke occurrence. Subsequently, multivariable logistic regression was subsequently applied to identify predictors with independent associations. The analysis identified the CHA$_2$DS$_2$-VASc score as the only significant independent predictor (odds ratio: 1.25; 95\% CI: 1.05–1.5), irrespective of atrial fibrillation status. These findings support the use of CHA$_2$DS$_2$-VASc as a simple preoperative risk stratification tool in cardiac surgical populations, with logistic regression serving primarily as an explanatory modeling approach rather than a predictive framework.

Traditionally, statistical approaches such as logistic and Cox regression have been widely used to identify key clinical predictors of postoperative outcomes in elderly patients. While effective for explanatory analysis, these methods typically rely on a limited number of predefined variables and assume linear relationships, which may overlook complex, nonlinear interactions inherent in stroke risk within heterogeneous surgical populations. Moreover, commonly used scoring systems—such as the Charlson Comorbidity Index, CHADS$_2$, and CHA$_2$DS$_2$-VASc—are constrained in their capacity to incorporate high-dimensional or time-dependent clinical data, such as longitudinal laboratory values and physiologic monitoring trends. These limitations underscore the need for more flexible, data-driven methodologies capable of leveraging the full granularity of electronic health records. In response, machine learning–based approaches have gained increasing traction for their ability to model nonlinear interactions, automate feature selection, and improve predictive performance across diverse clinical environments.

In recent years, machine learning (ML) has gained increasing prominence in clinical diagnostics and risk stratification, offering flexible alternatives to traditional statistical methods. Mu et al.~(2025) developed a Random Forest model for mortality prediction in sepsis-associated acute respiratory distress syndrome (ARDS) patients, achieving an Area Under the Receiver Operating Characteristic curve (AUROC) of 0.8015 and identifying blood urea nitrogen and the Simplified Acute Physiology Score II (SAPS II) as top predictors~\cite{mu2025predicting}. Similarly, Chen et al.~(2025) implemented an XGBoost-based framework to forecast Intensive Care Unit (ICU) readmission in acute pancreatitis patients, achieving superior discrimination (AUROC: 0.862, 95\% CI: 0.800--0.920; Accuracy: 0.889) and improved interpretability via SHapley Additive exPlanations (SHAP) analysis, which highlighted platelet count and SpO\textsubscript{2} as key features~\cite{chen2025predicting}. These machine learning techniques excel in capturing complex, nonlinear associations and higher-order interactions often overlooked by traditional approaches. Among them, tree-based ensemble methods have consistently demonstrated robustness and clinical relevance. In particular, CatBoost has emerged as a state-of-the-art algorithm for structured medical data due to its (1) mitigation of target leakage through ordered boosting, (2) native support for categorical variables and missing values without extensive preprocessing, and (3) built-in regularization with interpretable feature importance, which enhances model robustness and clinical utility.  Its effectiveness has been demonstrated in recent studies: Zhao et al.~(2020) used CatBoost to predict sepsis-induced coagulopathy, achieving an AUROC of 0.869 and outperforming conventional clinical scores~\cite{zhao2020sic}; Si et al.~(2025) applied it to in-hospital mortality prediction after cardiac arrest, yielding AUROC values of 0.904 (training) and 0.868 (testing), with SHAP-based insights supporting clinical relevance~\cite{si2025retrospective}. These applications, while focused on distinct outcomes, underscore the utility of CatBoost in addressing key challenges that also characterize postoperative stroke prediction—namely, heterogeneous ICU populations, high-dimensional perioperative data, and the need for early, interpretable risk stratification.

This study introduces a set of methodological innovations specifically designed to improve both the predictive robustness and clinical relevance of machine learning models for identifying postoperative stroke risk factors in elderly patients admitted to the surgical intensive care unit (SICU).

\begin{itemize}

  \item \textbf{Model Interpretability and Clinical Integration:} To promote transparency and facilitate clinical translation, we applied SHAP analysis to assess the global importance of input features. Influential predictors—such as creatinine levels, systolic blood pressure (SBP), and comorbidity profiles—aligned with established medical understanding. These findings improve the model’s transparency and emphasize its clinical utility in supporting perioperative decision-making for elderly SICU patients.

  \item \textbf{Hybrid Feature Selection Strategy:} We developed a two-stage feature selection framework that integrates Random Forest–based importance ranking with Recursive Feature Elimination and Cross-Validation (RFECV). This sequential procedure systematically reduced the initial pool of 80 features to 20 clinically meaningful variables while preserving predictive performance. To ensure both statistical rigor and clinical interpretability, the final feature set was further reviewed and endorsed by domain experts.

  \item \textbf{Modular Preprocessing and Class Imbalance Mitigation:} A modular preprocessing pipeline was implemented to address missing data, variable encoding, and class imbalance in a leakage-resistant manner. Numerical variables were imputed using the iterative Singular Value Decomposition (SVD) and standardized via z-score transformation, while categorical variables were mode-imputed and one-hot encoded. The Adaptive Synthetic Sampling (ADASYN) algorithm was applied within training folds only, ensuring that synthetic oversampling and all preprocessing steps were confined to training data during cross-validation, thereby preserving model generalizability and preventing information leakage.

 \item \textbf{Model Selection and Evaluation:} To identify the optimal classifier, we systematically compared eight machine learning models using a unified suite of performance metrics—namely, the area under the receiver operating characteristic curve (AUROC), overall accuracy, sensitivity, specificity, and the predictive values for both positive (PPV) and negative (NPV) outcomes. CatBoost consistently demonstrated superior results, achieving an AUROC of 0.8868 (95\% CI: 0.8802--0.8937) alongside a classification accuracy of 91.12\%. These findings informed its selection as the final model for interpretation and future clinical integration.

\end{itemize}

\section*{Methods}

\subsection*{Data Source and Study Design}
This study draws on two publicly available, high-resolution critical care datasets—MIMIC-III and MIMIC-IV—curated by the Laboratory for Computational Physiology at the Massachusetts Institute of Technology in collaboration with Beth Israel Deaconess Medical Center, Boston, MA. MIMIC-III includes over 40{,}000 de-identified ICU admissions from 2001 to 2012~\cite{johnson2016mimic}, while MIMIC-IV expands this to more than 76{,}000 admissions spanning 2008 to 2019, offering improved Electronic Health Record (EHR) compatibility and clinical detail~\cite{johnson2021mimic}. Both datasets are fully de-identified in accordance with HIPAA guidelines and were accessed under approved credentialing.

By jointly utilizing MIMIC-III and MIMIC-IV, we constructed a temporally and demographically heterogeneous SICU cohort of elderly surgical patients, enhancing model generalizability. A unified data science pipeline was employed encompassing cohort selection, data preprocessing, domain-driven feature engineering, model training, and validation procedures, as illustrated in Algorithm~\ref{alg:stroke}.

\begin{algorithm}[H]
\caption{\textbf{Data Preparation and Cohort Selection Pipeline for Stroke Prediction}}
\label{alg:stroke}
\begin{algorithmic}[1]
\Require MIMIC-III and MIMIC-IV ICU patients admitted to SICU
\Ensure Binary outcome: stroke diagnosis ($1 = \text{stroke}$, $0 = \text{no stroke}$)

\State \textbf{Step 1: Patient Selection}
\State Select SICU admissions from MIMIC-III and MIMIC-IV
\State Include solely the first ICU stay for each patient
\State Exclude patients aged $\leq 55$ and ICU stay $<48$ hours
\State Merge filtered MIMIC-III and MIMIC-IV cohorts
\State Remove rows with > 30\% missingness
\State Label stroke status using ICD-9 codes

\State \textbf{Step 2: Data Preprocessing}
\State Remove features with > 20\% 
\ForAll{numerical variables}
    \State Compute minimum, maximum, and average within the initial 24-hour window
    \State Handle missing data via iterative SVD-based imputation
    \State Standardize continuous features using z-score normalization
\EndFor

\ForAll{categorical variables}
    \State Impute missing values with the most frequent category (mode)
    \State Apply one-hot encoding
\EndFor

\State \textbf{Step 3: Statistical Analysis}
\State Evaluate baseline heterogeneity by examining feature distributions across training vs.~test sets and stroke vs.~non-stroke subgroups. For continuous variables, apply  independent samples t-tests, while categorical features are analyzed using chi-squared tests.

\State \textbf{Step 4: Feature Selection}
\State Apply RFECV using Random Forest (Gini-based importance)
\State Identify intersection with SHAP-based top features
\State Retain 20 shared features for model training

\State \textbf{Step 5: Model Development}
\ForAll{models $\in$ \{LR, NB, SVM, KNN, GBDT, XGBoost, LightGBM, CatBoost\}}
    \State Construct pipeline: preprocessing $\rightarrow$ ADASYN $\rightarrow$ classifier
    \State 5-fold cross-validation on training data with20 selected features
    \State Evaluate: AUROC, accuracy, sensitivity, specificity, F1-score, PPV, NPV (95\% CI)
\EndFor

\State \textbf{Step 6: Model Hyperparameter Optimization}
\State Perform 5-fold GridSearchCV on the best-performing model (CatBoost), optimizing hyperparameters based on mean AUROC

\State \textbf{Step 7: Model Interpretation}
\State SHAP summary analysis on CatBoost
\State Ablation study: remove features $\rightarrow$ AUROC
\State Visualize: top-20 SHAP feature summary and ablation impact plot

\end{algorithmic}
\end{algorithm}

\subsection*{Patient Selection}
Patients were identified from both the MIMIC-III and MIMIC-IV databases, two large publicly available critical care datasets containing de-identified health-related data from ICU admissions. MIMIC-III covers ICU stays from 2001 to 2012, while MIMIC-IV contains data from 2008 to 2019. Data extraction was performed using structured SQL queries to retrieve relevant patient demographics, diagnoses, vital signs, laboratory tests, and admission details, ensuring consistent and reproducible cohort construction across both datasets.

Cohort construction was conducted separately for the MIMIC-III and MIMIC-IV databases using a unified pipeline:  
(1) SICU admissions were initially identified based on surgical unit designations, defined by `first\_careunit' values of `MICU', `SICU', or `TSICU' in MIMIC-III, and `Medical/Surgical Intensive Care Unit (MICU/SICU)', `Surgical Intensive Care Unit (SICU)', or `Trauma SICU (TSICU)' in MIMIC-IV, yielding 26,648 admissions from MIMIC-III and 30,098 from MIMIC-IV;  
(2) only the first ICU admission per patient was retained to ensure independence of observations;  
(3) admissions with a length of ICU stay less than 48 hours were excluded;  
(4) patients aged 55 years or younger at the time of SICU admission were excluded to focus on an older adult population;  
(5) Stroke status was determined using International Classification of Diseases, Ninth Revision (ICD-9) diagnosis codes. Patients were classified as stroke-positive if any of the following codes were present: 430, 431, 432, 433, 434, or 436. This coding-based definition is consistent with prior validation studies assessing the reliability of administrative data in identifying stroke events~\cite{thigpen2015validity}.

After applying these criteria, 10,072 eligible admissions were retained from MIMIC-III and 9,979 from MIMIC-IV. The two subsets were subsequently merged, yielding a combined cohort of 20,051 SICU admissions. Clinical features, including demographics, comorbidities, vital signs (e.g., heart rate, blood pressure), and laboratory results, were collected within the initial 24-hour ICU period. This timeframe aligns with standard clinical protocols for early risk assessment and facilitates prompt prediction of adverse events. Records exceeding 30\% missing data among key variables were excluded to maintain data integrity, yielding a final cohort of 19,085 patients.

\begin{figure}[H]
\centering
\includegraphics[width=0.85\linewidth]{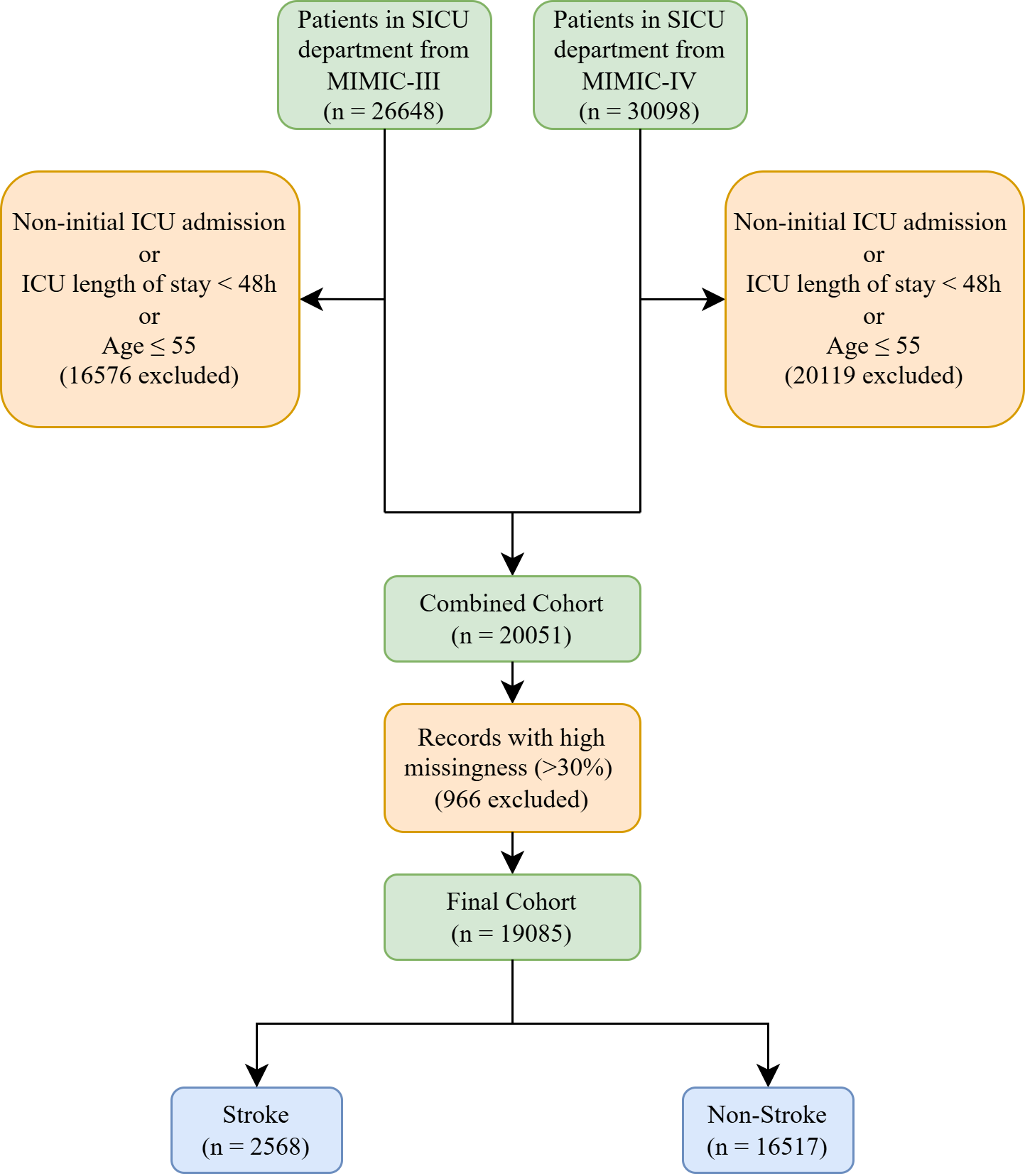}
\caption{\textbf{Patient Selection Process Flowchart Based on MIMIC-III and MIMIC-IV Databases}}
\label{fig:patient_selection}
\end{figure}

The detailed patient selection and data extraction process is depicted in Figure~\ref{fig:patient_selection}.

\subsection*{Data Preprocessing}
A comprehensive and systematic data preprocessing protocol was employed to guarantee data quality, consistency, and reliability before downstream analysis. In clinical datasets, preprocessing is particularly critical due to inherent challenges such as missing data, variable measurement units, and disparities in feature scales. Proper handling of these issues is essential to prevent biases, improve model robustness, and enhance interpretability, thereby ensuring that predictive models are both accurate and clinically meaningful.

Features with missing rates exceeding 20\% were excluded. Missing data in numerical variables were imputed using an iterative SVD algorithm~\cite{zhang2022postoperativestroke}, which leverages the low-rank structure commonly observed in clinical datasets to reconstruct incomplete entries. Unlike univariate approaches such as mean or median substitution, iterative SVD preserves multivariate correlations, thereby improving the accuracy and clinical plausibility of imputed values. This is particularly important in critical care settings, where physiological and laboratory parameters exhibit complex interdependencies essential for robust risk modeling.

After imputation, all continuous features were scaled via z-score normalization to harmonize measurement units and facilitate comparability. Specifically, to prevent information leakage and maintain the integrity of model evaluation, each feature was standardized using training-set-derived statistics—mean (\( \mu \)) and standard deviation (\( \sigma \))—to ensure inputs were centered and scaled consistently:
\begin{equation}
z_i = \frac{x_i - \mu}{\sigma}
\label{eq:zscore}
\end{equation}

 This normalization mitigates the influence of scale disparities, enhances numerical stability, and facilitates faster convergence in gradient-based optimization algorithms, ultimately improving model performance and interpretability~\cite{petersen2017normalization}.

Missing categorical values were imputed using mode substitution, a strategy that maintains the underlying distribution and mitigates potential imputation bias. One-hot encoding was then applied, with the first category dropped to prevent multicollinearity. This approach converts nominal variables into binary indicators while preserving feature matrix rank, ensuring stable coefficient estimation and interpretability. 
 
 Applying type-specific imputation strategies improves data integrity and ensures a coherent feature representation. Collectively, these preprocessing steps facilitate robust model training, enhance the performance of downstream machine learning algorithms, and provide a stable foundation for clinically meaningful and generalizable predictive modeling.

\begin{figure}[H]
\centering
\includegraphics[width=0.7\linewidth]{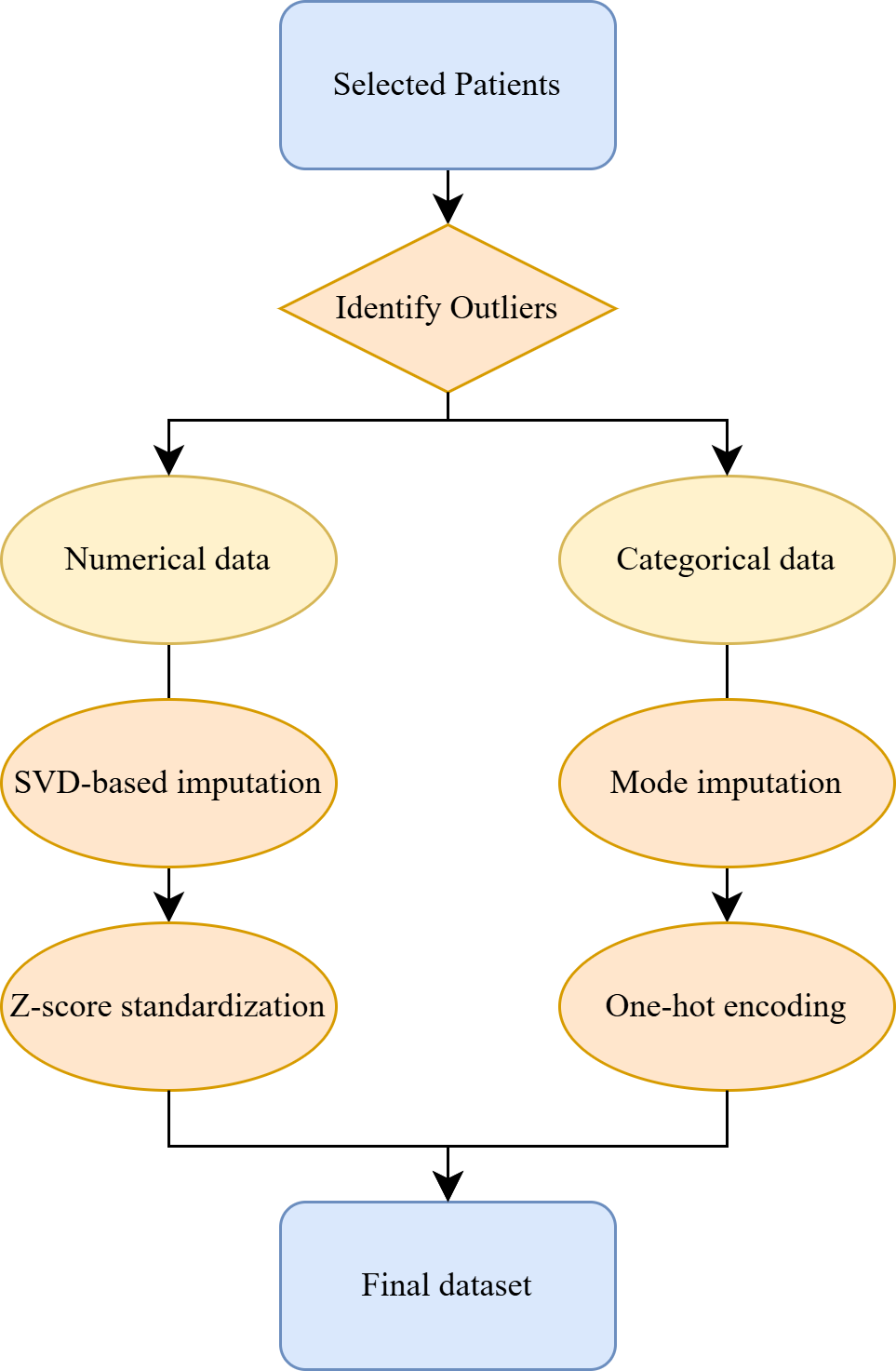}
\caption{\textbf{Data Preprocessing and Feature Selection Pipeline Flowchart}}
\label{fig:data_preprocessing_pipeline}
\end{figure}

An overview of the full data preprocessing pipeline is shown in Figure~\ref{fig:data_preprocessing_pipeline}.

\subsection*{Statistical Analysis}
Baseline characteristics were compared across both the training and test sets, as well as between patients with and without postoperative stroke, to assess potential distributional differences. For continuous variables—including vital signs and laboratory values—two-sample Student’s \( t \)-tests were employed under the assumption of approximate normality; this approach is generally robust to mild deviations from normality~\cite{lumley2002importance}. For categorical variables such as comorbidities and demographic features, Pearson’s chi-square tests were used to evaluate group-wise differences in frequency distributions~\cite{rana2015chi}. Test selection was guided by conventional statistical assumptions regarding variable type and sample size. A two-sided \( p \)-value of less than 0.05 was considered indicative of statistical significance. All tests were conducted independently for descriptive comparison purposes only and were not used to inform feature selection or model training.

\subsection*{Feature Selection}
 To derive a clinically meaningful and statistically sound feature set, candidate predictors were selected based on prior studies in stroke and critical care outcome modeling~\cite{zhang2022postoperativestroke, heo2019stroke, sirsat2020machine, wu2020stroke, dunham2017perioperative, bolourani2021machine}, complemented by input from clinical experts. For each continuous physiological and laboratory measure, the minimum, maximum, and mean values during the initial 24-hour ICU period were calculated, resulting in three features per variable. The final feature set comprised 80 variables grouped into four major categories:

\begin{itemize}
\item \textbf{Demographics and Administrative Indicators:} This group included age, sex, insurance type, marital status, and ethnicity, capturing baseline patient characteristics known to influence disease trajectory and healthcare access disparities.

\item \textbf{Vital signs:} Acute physiological status was assessed using vital signs recorded during the initial 24 hours in the ICU, including heart rate, systolic and diastolic blood pressure (SBP, DBP), respiratory rate, peripheral oxygen saturation (SpO\textsubscript{2}), and body temperature.

\item \textbf{Laboratory results:} A comprehensive panel of laboratory tests was extracted to evaluate organ function, coagulation status, oxygenation, and metabolic response. Renal markers included blood urea nitrogen (BUN) and creatinine. Coagulation indices comprised international normalized ratio (INR), prothrombin time (PT), and partial thromboplastin time (PTT), reflecting coagulation abnormalities. Key metabolic and inflammatory indicators encompassed serum glucose, anion gap, lactate, potassium, and white blood cell count (WBC). Liver function was represented by albumin and total bilirubin, while hematologic status was assessed using hematocrit and hemoglobin. For each variable, descriptive statistics (minimum, maximum, and mean) were computed based on measurements documented during the first 24-hour window following ICU admission~\cite{dunham2017perioperative, wu2020stroke, bolourani2021machine, zhang2022postoperativestroke}.

\item \textbf{Comorbidities:} Chronic conditions with established relevance to stroke outcomes were incorporated, including peripheral vascular disease, hypertension, chronic pulmonary disease, diabetes, renal disease, liver disease, severe liver disease, peptic ulcer disease, sepsis, congestive heart failure, ischemic heart disease, cerebrovascular disease, cancer, dementia, and rheumatic disease \cite{ wu2020stroke, dunham2017perioperative}.
\end{itemize}

 To enhance data integrity and support robust model development, variables exhibiting over 20\% missingness across the integrated MIMIC-III and MIMIC-IV cohorts were removed during the preprocessing stage. This resulted in the removal of over 30 derived features, including clinically relevant but sparsely documented variables such as mean albumin, minimum lactate, and maximum bilirubin, along with various SpO\textsubscript{2}, glucose, and white blood cell statistics.

 To refine the remaining feature set, we employed a hybrid selection strategy combining model-driven and explanation-based approaches to improve both performance and interpretability. RFECV was performed using a Random Forest classifier, selected for its robustness to feature scaling, tolerance of multicollinearity, and ability to quantify feature importance. The RFECV procedure iteratively removed low-importance variables while optimizing five-fold cross-validated performance. Feature importance was measured by summing the total reduction in Gini impurity attributed to each feature across all nodes in the ensemble, formally defined as:
\begin{equation}
I(x_i) = \sum_{t \in T} p(t) \cdot \Delta_i(t) \cdot f(t),
\end{equation}
where \( I(x_i) \) quantifies the contribution of feature \( x_i \) across the forest; \( T \) denotes the set of all tree nodes; \( p(t) \) is the fraction of instances routed through node \( t \); \( \Delta_i(t) \) indicates the reduction in impurity attributed to \( x_i \) at node \( t \); and \( f(t) \) reflects how often \( x_i \) is selected for splitting at that node.

In parallel, global SHAP values~\cite{lundberg2017unified} were computed to quantify each feature’s average marginal contribution to model predictions. SHAP, grounded in cooperative game theory, provides locally accurate, additive explanations that complement tree-based importance scores.

The final feature subset was identified by intersecting the top-ranked variables selected from both Recursive Feature Elimination with Cross-Validation (RFECV) and SHAP value analyses. This combined approach ensured a balance between statistical significance and clinical interpretability, resulting in a concise and transparent set of 20 features for subsequent model development.

These 20 features spanned key clinical domains, encompassing demographics, vital signs, laboratory measurements, and chronic comorbidities. Specifically, the set included age; vital signs such as respiratory rate, heart rate, SBP and DBP, and body temperature; laboratory results including creatinine, BUN, and PT; and comorbidities such as hypertension, heart failure, renal disease, chronic pulmonary disease, sepsis, ischemic heart disease, and cerebrovascular disease. To capture dynamic physiological changes, continuous variables were represented not only by measures of central tendency (e.g., mean) but also by range statistics (minimum and maximum values). Collectively, these features encapsulated clinically meaningful signals critical for accurately predicting postoperative stroke risk among critically ill elderly patients.

These 20 features, shown in Table~\ref{tab:final_features}, encompass key demographics, vital signs, laboratory results, and comorbidities. Specifically, the set included age; vital signs such as respiratory rate, heart rate, SBP and DBP, and body temperature; laboratory results including creatinine, BUN, and PT; and comorbidities such as hypertension, heart failure, renal disease, chronic pulmonary disease, sepsis, ischemic heart disease, and cerebrovascular disease. To capture dynamic physiological changes, continuous variables were represented not only by measures of central tendency (e.g., mean) but also by range statistics (minimum and maximum values). Collectively, these features encapsulated clinically meaningful signals critical for accurately predicting postoperative stroke risk among critically ill elderly patients.

\begin{table}[H]
\noindent
\caption{\textbf{Selected Feature Set for Postoperative Stroke Prediction in Elderly SICU Patients}}
\label{tab:final_features}
\small
\renewcommand{\arraystretch}{1.1}
\rowcolors{2}{white}{white}
\begin{tabularx}{\textwidth}{l|>{\raggedright\arraybackslash}X}
\hline
\rowcolor[HTML]{D9EAD3}
\textbf{Category} & \textbf{Features} \\
\hline
Demographics & Age \\
\hline
Vital Signs & Respiratory rate , Heart rate, SBP, DBP, Temperature \\
\hline
Laboratory Results & Creatinine, BUN, PT \\
\hline
Comorbidities & Hypertension, Heart failure, Renal disease, Chronic pulmonary disease, Sepsis, Ischemic heart disease, Cerebrovascular disease \\
\hline
\end{tabularx}
\end{table}

\subsection*{Model Development and Evaluation}
For the purpose of predicting postoperative stroke in elderly SICU patients, a supervised learning model was constructed based on 20 selected clinical, physiological, and laboratory features. The combined dataset from MIMIC-III and MIMIC-IV was evaluated using stratified 5-fold cross-validation to preserve the outcome distribution and mitigate sampling variability.

\begin{figure}[H]
    \centering
    \includegraphics[width=0.95\linewidth]{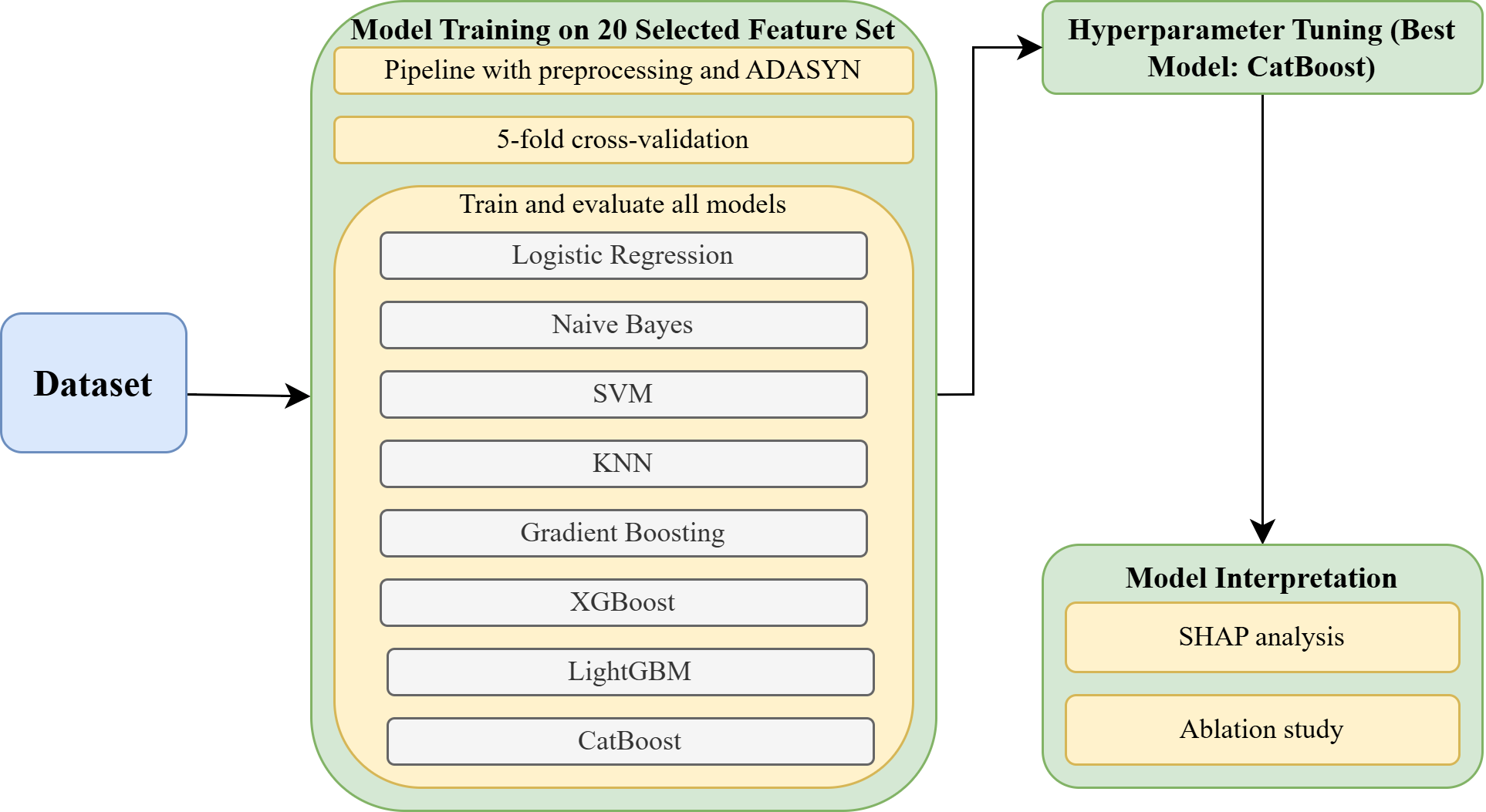}
    \caption{\textbf{Model Training and Evaluation Workflow Flowchart Detailing Embedded Feature Selection, Final Model Optimization, and Interpretation}}
    \label{fig:modeling_feature_selection}
\end{figure}

A set of eight diverse machine learning algorithms, selected for their varied methodologies and practical strengths, were implemented and assessed: Logistic Regression (LR), Naive Bayes (NB), Support Vector Machine (SVM), K-Nearest Neighbors (KNN),Gradient Boosting Machines (GBM), XGBoost, LightGBM, and CatBoost. For each model, a standardized pipeline was constructed comprising feature preprocessing, class imbalance correction using the ADASYN oversampling algorithm~\cite{he2008adasyn}, and model training. Model performance was assessed through stratified 5-fold cross-validation, evaluating metrics such as AUROC, accuracy, sensitivity, specificity, F1-score, positive predictive value (PPV), and negative predictive value (NPV). Confidence intervals at the 95\% level were derived using bootstrap resampling. The comprehensive modeling pipeline is depicted in Figure~\ref{fig:modeling_feature_selection}.

Each model was chosen to reflect complementary algorithmic strengths and to capture the diverse patterns inherent in heterogeneous clinical data. LR was included as a baseline due to its simplicity, interpretability, and established use in clinical research. Its linear nature and probabilistic output make it particularly useful for risk stratification tasks. NB was incorporated for its computational efficiency and strong performance on high-dimensional, low-sample-size datasets, despite its conditional independence assumption.

SVM was chosen for its ability to construct high-margin classifiers and capture complex boundaries in feature space, especially when combined with kernel functions. KNN, while simple in design, was included for its instance-based learning approach and capacity to model local data structures without prior assumptions about feature distributions.

We also included three gradient boosting algorithms—GBM, XGBoost, and LightGBM—each known for their strong empirical performance. GBM provides a flexible boosting framework and served as a traditional benchmark. XGBoost improves upon standard boosting with regularized loss functions and second-order gradient optimization, making it adept at capturing subtle feature interactions and controlling overfitting. LightGBM leverages histogram-based feature binning and a leaf-wise growth strategy to accelerate training while maintaining strong predictive performance, particularly in sparse or high-dimensional datasets.

CatBoost was additionally incorporated for its advanced handling of categorical variables through ordered boosting and its resistance to overfitting in small to medium-sized datasets. This characteristic made it especially well-suited for our clinical setting, which often involves variable dependencies, missing data, and nonlinear relationships. 

Following the initial comparison, we optimized the hyperparameters of the best model, CatBoost, using a 5-fold cross-validated grid search. Key hyperparameters such as learning rate, tree depth, regularization, and feature sampling ratios were tuned to optimize the mean AUROC, which was used as the main criterion for model selection.

Final model performance was evaluated using AUROC, with 95\% confidence intervals estimated from 2{,}000 bootstrap replicates to quantify statistical uncertainty and generalizability.Let \(\hat{y}_i\) be the predicted probability for sample \(i\) and \(y_i\) the true label. CatBoost was trained by minimizing the following regularized objective:
\begin{equation}
    \mathcal{L}(\theta) = \sum_{i=1}^{n} \ell(y_i, \hat{y}_i) + \Omega(f),
\end{equation}
where \(\ell\) is the binary cross-entropy loss function, and \(\Omega(f)\) is a penalty term to control model complexity and prevent overfitting. This approach aligns with common gradient-boosted tree methodologies and promotes both accuracy and robustness in complex clinical datasets.

The systematic modeling framework enabled consistent classifier comparison, effectively managed data imbalance and noise, and delivered a robust, interpretable model for stratifying postoperative stroke risk in elderly SICU patients.

\subsection*{Model Interpretability and Feature Contribution Analysis}

To enhance interpretability and ensure clinical plausibility, we conducted a two-pronged feature evaluation strategy comprising SHAP-based attribution analysis and iterative ablation testing.

First, we utilized SHAP~\cite{lundberg2017unified} to evaluate the contribution of individual features to the model’s predictions. Grounded in cooperative game theory, SHAP values provide additive attributions quantifying both the direction and strength of each feature’s influence. Mathematically, the SHAP value for feature \(i\) is expressed as:
\begin{equation}
\phi_i = \sum_{S \subseteq N \setminus \{i\}} \frac{|S|! (|N|-|S|-1)!}{|N|!} \left[ f(S \cup \{i\}) - f(S) \right],
\end{equation}
where \(\phi_i\) represents the average marginal effect of adding feature \(i\) across all subsets \(S\) that exclude it. Global feature importance was visualized using SHAP summary plots, highlighting the top 20 features by mean absolute SHAP value. These findings aligned well with established clinical knowledge, emphasizing key predictors including age, creatinine, SBP, and prothrombin time.

Second, we conducted an ablation study to evaluate the contribution and stability of each feature. Beginning with the full CatBoost model incorporating all 20 features, we sequentially excluded one feature per iteration, retraining the model to measure the resulting change in performance. The difference in AUROC after excluding feature \(i\) was calculated as:
\begin{equation}
\Delta_i = \text{AUROC}(f_{\text{full}}) - \text{AUROC}(f_{-i}),
\end{equation}
where \(f_{\text{full}}\) represents the model trained with the complete feature set, and \(f_{-i}\) is the model trained without feature \(i\). Features whose removal led to AUROC improvement were permanently dropped, with the refined model serving as the basis for subsequent iterations. This iterative pruning proceeded until no further performance gains were achieved. The ablation outcomes were depicted through impact plots, providing clear visualization of feature importance for classification accuracy.

Together, the SHAP evaluation and systematic feature ablation enhanced the interpretability of the final model while confirming the essential role of key clinical predictors. These explainability methods bolster model transparency and reinforce its suitability for application in critical clinical decision-making settings.

\section*{Results}
\subsection*{Study Cohort Profile and Comparative Statistics}
The study cohort consisted of 19{,}085 elderly patients admitted to the SICU in the postoperative period, assessed for the occurrence of in-hospital stroke. Among them, 2{,}568 patients (13.5\%) developed postoperative stroke (Group 1), while the remaining 16{,}517 (86.5\%) did not (Group 0). To obtain robust estimates of model performance, we employed stratified 5-fold cross-validation, preserving the distribution of outcome classes within each fold during both training and validation.

The primary aim of this statistical analysis was to assess whether significant differences existed between cohorts—both in terms of model development splits and stroke outcomes—across the full set of clinical features. Establishing such equivalence is essential to ensure that predictive models are generalizable and unbiased. Table~\ref{tab:train_test_stats} presents a comparison of continuous variables between training and test sets using independent two-sample t-tests. The distributions of key variables such as age, heart rate, creatinine, and SBP did not differ significantly between groups ($p > 0.05$), indicating well-balanced cohort construction. A few features—such as mean and maximum diastolic blood pressure—did exhibit modest but statistically significant variation, suggesting minor shifts in baseline hemodynamic status.

Table~\ref{tab:stroke_comparison} further compares continuous features between stroke and non-stroke groups. Stroke patients had significantly lower creatinine, lower heart rate , and higher SBP, all with $p < 0.0001$. Coagulation-related features such as PT and PTT were also markedly different, potentially reflecting systemic inflammation or vascular instability. These findings support the physiologic plausibility of selected predictors and reinforce their relevance in the pathophysiology of postoperative stroke.

In parallel, chi-square testing was applied to assess distributional variation in categorical variables, including comorbidities and sociodemographic factors. However, due to pronounced class imbalance—common in real-world critical care datasets—many variables lacked sufficient variability to meet the assumptions of chi-square testing. For example, over 90\% of patients in variables such as dementia, cerebrovascular disease, severe liver disease, and cancer belonged to a single category, precluding valid statistical comparison. While formal evaluation was limited, the absence of large detectable shifts in these categorical variables further supports the consistency and representativeness of the dataset.

By contrast, comparisons between stroke and non-stroke populations yielded multiple statistically significant differences across categorical features (Table~\ref{tab:chi2_stroke}). Conditions such as hypertension, diabetes, ischemic heart disease, chronic pulmonary disease, renal disease, and sepsis were notably more prevalent among stroke patients ($p < 0.05$), in agreement with existing literature on stroke pathogenesis in surgical and ICU populations. Furthermore, sociodemographic factors such as race and insurance type also differed significantly between groups, highlighting potential disparities in care or baseline risk exposure. These findings emphasize the multifactorial nature of postoperative stroke and underscore the clinical relevance of incorporating both physiologic and categorical variables in predictive modeling.

\begin{table}[H]
\centering
\caption{\textbf{Comparison of Numerical Feature Distributions in Training and Test Sets via T-Test}}
\label{tab:train_test_stats}
\small
\renewcommand{\arraystretch}{1.1}
\rowcolors{2}{white}{white}
\begin{tabularx}{\textwidth}{l|c|c|c}
\hline
\rowcolor[HTML]{D9EAD3}
\textbf{Feature} & \textbf{Training Set (Mean ± SD)} & \textbf{Test Set (Mean ± SD)} & \textbf{P-value} \\
\hline
dbp\_mean         & 64.05 ± 45.68    & 67.00 ± 101.23   & 0.0074 \\
dbp\_max          & 119.38 ± 10.66   & 183.30 ± 23.84   & 0.0114 \\
temp\_mean        & 98.34 ± 3.18     & 98.75 ± 26.43    & 0.0606 \\
temp\_max         & 100.04 ± 17.62   & 102.47 ± 15.23   & 0.0639 \\
ptt\_mean         & 35.97 ± 16.26    & 36.46 ± 16.91    & 0.0981 \\
hr\_mean          & 86.55 ± 16.32    & 86.07 ± 16.27    & 0.1041 \\
rr\_mean          & 19.45 ± 4.17     & 19.34 ± 4.07     & 0.1434 \\
ptt\_max          & 40.28 ± 24.46    & 40.89 ± 25.01    & 0.1724 \\
ptt\_min          & 32.38 ± 12.05    & 32.68 ± 12.56    & 0.1802 \\
age               & 80.49 ± 43.88    & 81.38 ± 46.05    & 0.2633 \\
pt\_min           & 15.28 ± 5.37     & 15.18 ± 4.96     & 0.2938 \\
hr\_max           & 107.13 ± 54.85   & 106.39 ± 22.24   & 0.4139 \\
sbp\_mean         & 118.25 ± 16.89   & 118.01 ± 14.85   & 0.4224 \\
rr\_max           & 28.39 ± 26.11    & 28.05 ± 7.17     & 0.4239 \\
sbp\_max          & 147.38 ± 13.54   & 145.85 ± 25.89   & 0.4761 \\
bun\_max          & 21.02 ± 12.60    & 19.03 ± 6.71     & 0.4797 \\
bun\_min          & 18.04 ± 10.42    & 16.06 ± 5.78     & 0.5275 \\
pt\_mean          & 15.92 ± 6.09     & 15.87 ± 5.94     & 0.6823 \\
creatinine\_mean  & 1.42 ± 1.36      & 1.43 ± 1.33      & 0.7744 \\
creatinine\_max   & 1.51 ± 1.47      & 1.52 ± 1.42      & 0.7799 \\
creatinine\_min   & 1.33 ± 1.26      & 1.33 ± 1.25      & 0.8136 \\
pt\_max           & 16.69 ± 7.67     & 16.71 ± 8.46     & 0.8795 \\
\hline
\end{tabularx}
\end{table}

\begin{table}[H]
\centering
\caption{\textbf{Comparison of Numerical Feature Distributions in Stroke and Non-Stroke Patient Groups via T-Test}}
\label{tab:stroke_comparison}
\small
\renewcommand{\arraystretch}{1.1}
\rowcolors{2}{white}{white}
\begin{tabularx}{\textwidth}{l|c|c|c}
\hline
\rowcolor[HTML]{D9EAD3}
\textbf{Feature} & \textbf{Non-Stroke (Mean ± SD)} & \textbf{Stroke (Mean ± SD)} & \textbf{P-value} \\
\hline
ptt\_max          & 40.91 ± 24.70  & 37.13 ± 23.49  & <0.0001 \\
pt\_min           & 15.46 ± 5.50   & 14.03 ± 3.46   & <0.0001 \\
bun\_min          & 27.67 ± 20.17  & 21.45 ± 14.61  & <0.0001 \\
ptt\_min          & 32.82 ± 12.44  & 30.01 ± 9.81   & <0.0001 \\
hr\_max           & 107.77 ± 23.06 & 101.90 ± 23.20 & <0.0001 \\
sbp\_max          & 145.30 ± 69.73 & 158.48 ± 69.21 & <0.0001 \\
creatinine\_mean  & 1.46 ± 1.35    & 1.16 ± 1.32    & <0.0001 \\
creatinine\_max   & 1.56 ± 1.46    & 1.21 ± 1.41    & <0.0001 \\
rr\_mean          & 19.56 ± 4.23   & 18.56 ± 3.46   & <0.0001 \\
pt\_max           & 16.96 ± 8.18   & 14.96 ± 4.72   & <0.0001 \\
hr\_mean          & 87.45 ± 16.34  & 80.09 ± 14.59  & <0.0001 \\
rr\_max           & 28.61 ± 25.20  & 26.44 ± 6.27   & <0.0001 \\
sbp\_mean         & 116.94 ± 15.53 & 126.28 ± 19.91 & <0.0001 \\
ptt\_mean         & 36.50 ± 16.57  & 33.31 ± 14.90  & <0.0001 \\
pt\_mean          & 16.13 ± 6.31   & 14.46 ± 3.86   & <0.0001 \\
creatinine\_min   & 1.36 ± 1.26    & 1.11 ± 1.19    & <0.0001 \\
dbp\_mean         & 64.06 ± 29.70  & 68.39 ± 18.37  & 0.0008 \\
dbp\_max          & 126.54 ± 24.14 & 125.32 ± 17.40 & 0.1584 \\
bun\_max          & 20.14 ± 11.86  & 21.73 ± 12.27  & 0.5707 \\
temp\_max         & 100.62 ± 77.93 & 99.90 ± 1.70   & 0.6420 \\
age               & 80.64 ± 44.37  & 80.84 ± 44.00  & 0.8321 \\
temp\_mean        & 98.43 ± 13.04  & 98.38 ± 2.29   & 0.8369 \\
\hline
\end{tabularx}
\end{table}

\begin{table}[H]
\centering
\caption{\textbf{Comparison of Categorical Feature Distributions in Stroke and Non-Stroke Patient Groups via Chi-square Test}}
\label{tab:chi2_stroke}
\small
\renewcommand{\arraystretch}{1.1}
\rowcolors{2}{white}{white}
\begin{tabularx}{0.6\textwidth}{l|c}
\hline
\rowcolor[HTML]{D9EAD3}
\textbf{Feature} & \textbf{P-value} \\
\hline
severe\_liver\_disease               & <0.0001 \\
chronic\_pulmonary\_disease\_strict & 0.0853 \\
rheumatic\_disease                   & <0.0001 \\
cerebrovascular\_disease             & <0.0001 \\
dementia                             & Skipped (low variability) \\
peptic\_ulcer\_disease               & 0.1537 \\
hypertension                        & <0.0001 \\
ischemic\_heart\_disease             & <0.0001 \\
liver\_disease                     & <0.0001 \\
renal\_disease                     & <0.0001 \\
congestive\_heart\_failure           & <0.0001 \\
insurance                          & 0.0087 \\
sepsis                             & <0.0001 \\
heart\_failure                     & <0.0001 \\
race                               & 0.0011 \\
chronic\_pulmonary\_disease          & <0.0001 \\
diabetes                           & <0.0001 \\
gender                             & 0.3433 \\
cancer                             & 0.2504 \\
peripheral\_vascular\_disease        & 0.1587 \\
\hline
\end{tabularx}
\end{table}

\subsection*{Evaluation and Comparison of Model Performance}
\begin{figure}[H]
\centering
\includegraphics[width=0.9\textwidth]{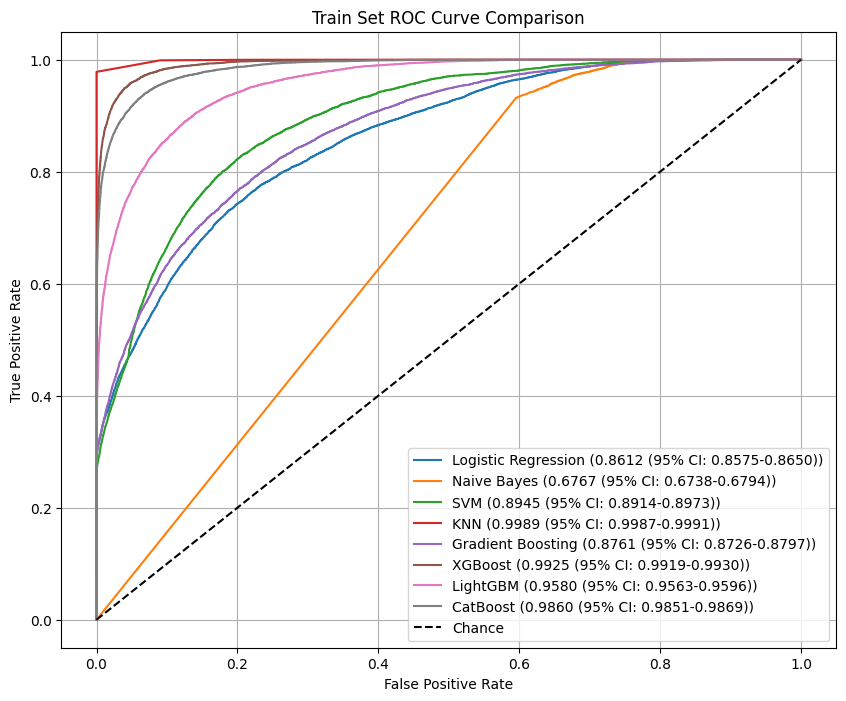}
\caption{\textbf{ROC curve for the training cohort}}
\label{fig:roc_train}
\end{figure}

\begin{figure}[H]
\centering
\includegraphics[width=0.9\textwidth]{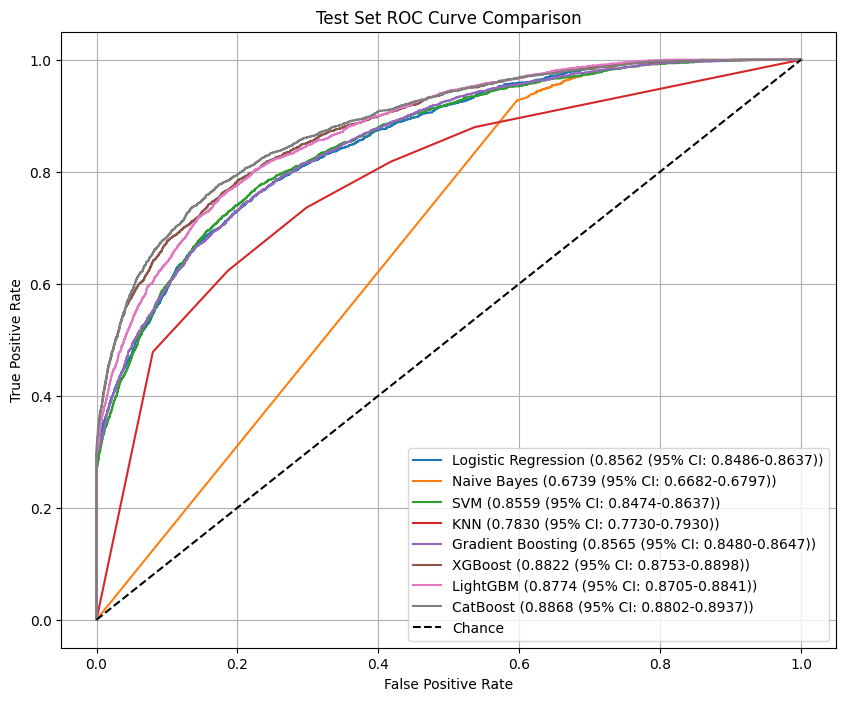}
\caption{\textbf{ROC curve for the test cohort}}
\label{fig:roc_test}
\end{figure}

To comprehensively evaluate the predictive performance and robustness of eight machine learning classifiers for postoperative stroke risk in elderly SICU patients, we performed a comparative analysis. Model performance was evaluated through AUROC, accuracy, sensitivity, specificity, PPV, NPV, and F1 score, each accompanied by 95\% confidence intervals. These metrics were derived using stratified 5-fold cross-validation on the training dataset and subsequently validated on an independent test cohort. Figure~\ref{fig:roc_train} and Figure~\ref{fig:roc_test} illustrate the Receiver Operating Characteristic (ROC) curves for the training and testing cohorts, respectively. The corresponding quantitative performance metrics are summarized in Table~\ref{tab:train_metrics} and Table~\ref{tab:test_metrics}.

Among the training dataset, ensemble methods—especially XGBoost, CatBoost, and LightGBM—demonstrated the best overall predictive performance. XGBoost obtained the highest AUROC (0.9925) and F1 score (0.8818), while CatBoost attained the highest specificity (0.9963), PPV (0.9681), and accuracy (0.9599), reflecting both strong discrimination and well-calibrated predictions. LightGBM also performed competitively (AUROC: 0.9580). KNN achieved near-perfect sensitivity (0.9999) and NPV (1.0000), although its lower PPV (0.4194) may indicate overfitting to the training data. In contrast, LR and NB exhibited weaker discriminative performance, particularly in PPV and F1 score, underscoring their limitations in modeling nonlinear and high-dimensional clinical patterns.

To assess real-world generalizability beyond cross-validated training performance, we further evaluated each model on an independent test set. CatBoost achieved the highest AUROC of 0.8868 (95\% CI: 0.8802–0.8937), accuracy of 0.9112, and specificity of 0.9817, along with a strong F1 score of 0.5817 and PPV of 0.7957, demonstrating the best overall performance among the evaluated models. XGBoost and LightGBM also showed high generalization capacity (AUROCs: 0.8822 and 0.8774, respectively). Interestingly, LR maintained a competitive AUROC (0.8562) and high sensitivity (0.8071), supporting its continued relevance in some clinical applications. In contrast, KNN and NB displayed substantially reduced test-time precision (PPV: 0.2775 and 0.1815, respectively), raising concerns about their suitability for high-stakes clinical prediction in postoperative stroke identification.

The superior test performance of gradient boosting models—CatBoost, LightGBM, and XGBoost—can be attributed to their capacity to capture nonlinear feature interactions, handle heterogeneous data types, and natively accommodate missingness. These structural advantages, combined with robustness to multicollinearity and class imbalance, make them well suited for complex ICU datasets. Among them, CatBoost's use of ordered boosting and native categorical handling likely contributed to its improved precision and calibration under class-imbalanced conditions, enabling better identification of true positives without sacrificing specificity.

Clinically, CatBoost’s high specificity (0.9817) and NPV (0.9209) are particularly valuable for ruling out low-risk patients and minimizing unnecessary interventions. Although its sensitivity (0.4585) was moderate, the strong PPV (0.7957) ensures that most flagged high-risk cases are credible, optimizing resource allocation in high-stakes ICU environments. This trade-off—prioritizing reliable alerts over exhaustive detection—is well aligned with real-world constraints in postoperative stroke management.

CatBoost stood out as the most suitable model for clinical deployment, based on its overall effectiveness. Its ability to maintain strong performance across key metrics, combined with interpretability and stability demonstrated in subsequent analyses, supports its integration into decision-support workflows for early risk stratification in elderly postoperative ICU patients.

\begin{table}[H]
\small
\renewcommand{\arraystretch}{1.2}
\centering
\caption{\textbf{Training Set Performance Comparison of Classifiers}}
\label{tab:train_metrics}
\begin{adjustbox}{width=\textwidth}
\begin{tabular}{l|c|c|c|c|c|c|c}
\hline
\rowcolor[HTML]{D9EAD3}
\textbf{Model} & \textbf{AUROC} & \textbf{Accuracy} & \textbf{F1 Score} & \textbf{Sensitivity} & \textbf{Specificity} & \textbf{PPV} & \textbf{NPV} \\
\hline
Logistic Regression & 0.8612 & 0.7272 & 0.4450 & 0.8125 & 0.7140 & 0.3064 & 0.9608 \\
\hline
Naive Bayes         & 0.6767 & 0.4100 & 0.3072 & 0.9724 & 0.3225 & 0.1824 & 0.9869 \\
\hline
SVM                 & 0.8945 & 0.7657 & 0.4975 & 0.8620 & 0.7507 & 0.3497 & 0.9722 \\
\hline
KNN                 & 0.9989 & 0.8138 & 0.5910 & 0.9999 & 0.7848 & 0.4194 & 1.0000 \\
\hline
Gradient Boosting   & 0.8761 & 0.8579 & 0.5524 & 0.6515 & 0.8900 & 0.4794 & 0.9426 \\
\hline
XGBoost             & 0.9925 & 0.9706 & 0.8818 & 0.8155 & 0.9947 & 0.9598 & 0.9720 \\
\hline
LightGBM            & 0.9580 & 0.9359 & 0.7194 & 0.6108 & 0.9864 & 0.8749 & 0.9422 \\
\hline
CatBoost            & 0.9860 & 0.9599 & 0.8294 & 0.7256 & 0.9963 & 0.9681 & 0.9589 \\
\hline
\end{tabular}
\end{adjustbox}
\end{table}

\begin{table}[H]
\small
\renewcommand{\arraystretch}{1.2}
\centering
\caption{\textbf{Test Set Performance Comparison of Classifiers}}
\label{tab:test_metrics}
\begin{adjustbox}{width=\textwidth}
\begin{tabular}{l|c|c|c|c|c|c|c}
\hline
\rowcolor[HTML]{D9EAD3}
\textbf{Model} & \textbf{AUROC} & \textbf{Accuracy} & \textbf{F1 Score} & \textbf{Sensitivity} & \textbf{Specificity} & \textbf{PPV} & \textbf{NPV} \\
\hline
Logistic Regression & 0.8562 & 0.7249 & 0.4412 & 0.8071 & 0.7121 & 0.3036 & 0.9596 \\
\hline
Naive Bayes         & 0.6739 & 0.4089 & 0.3057 & 0.9676 & 0.3221 & 0.1815 & 0.9846 \\
\hline
SVM                 & 0.8559 & 0.7480 & 0.4590 & 0.7946 & 0.7407 & 0.3228 & 0.9587 \\
\hline
KNN                 & 0.7830 & 0.7065 & 0.4030 & 0.7361 & 0.7019 & 0.2775 & 0.9448 \\
\hline
Gradient Boosting   & 0.8565 & 0.8493 & 0.5267 & 0.6231 & 0.8845 & 0.4563 & 0.9378 \\
\hline
XGBoost             & 0.8822 & 0.9093 & 0.5963 & 0.4979 & 0.9733 & 0.7433 & 0.9257 \\
\hline
LightGBM            & 0.8774 & 0.9029 & 0.5568 & 0.4534 & 0.9728 & 0.7216 & 0.9196 \\
\hline
\rowcolor[HTML]{FDE9D9}
CatBoost            & 0.8868 & 0.9112 & 0.5817 & 0.4585 & 0.9817 & 0.7957 & 0.9209 \\
\hline
\end{tabular}
\end{adjustbox}
\end{table}

\subsection*{Model Explanation and Clinical Insight via SHAP}
\begin{figure}[H]
\centering
\includegraphics[width=0.7\textwidth]{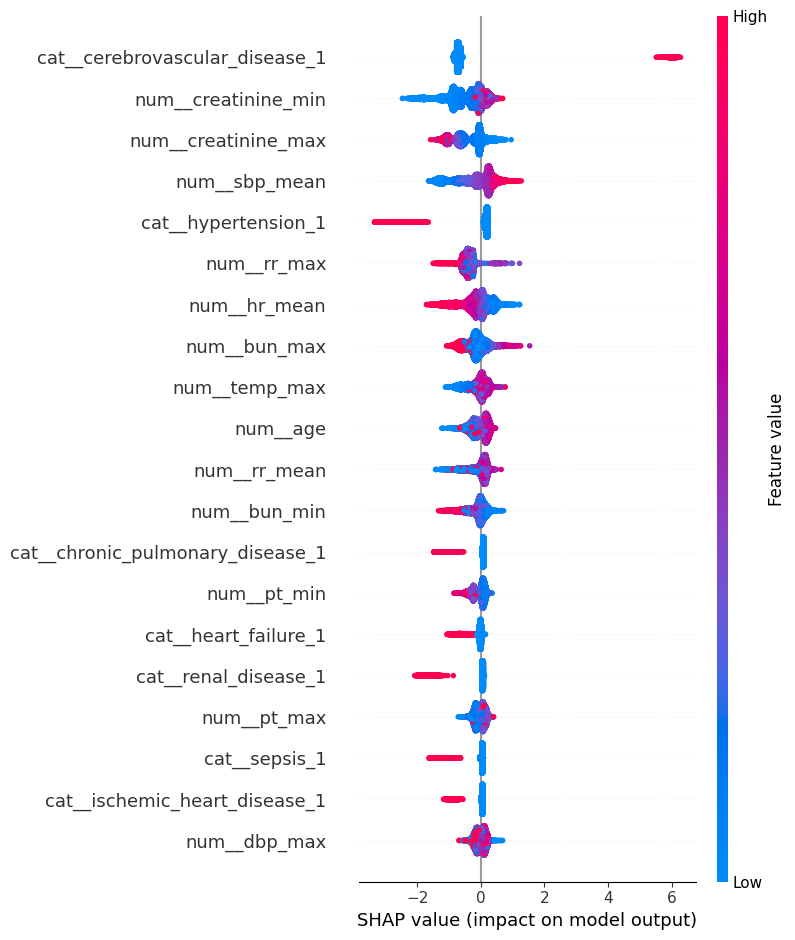}
\caption{\textbf{SHAP Summary Plot Showing Feature Contributions and Interpretability in the CatBoost Model}}
\label{fig:shap_summary}
\end{figure}

To support clinical interpretability, SHAP values were utilized to measure the individual contribution of each feature to the CatBoost model’s postoperative stroke risk predictions. Figure~\ref{fig:shap_summary} presents the global SHAP summary plot, ranking the top 20 predictors by mean absolute SHAP value. The horizontal axis represents both the strength and direction of each feature’s impact on the model predictions, while the color gradient indicates the original feature values (red for high, blue for low).

Cerebrovascular disease history emerged as the most influential predictor, with its presence substantially increasing predicted stroke risk—consistent with established clinical knowledge of recurrence susceptibility. Elevated creatinine levels (both minimum and maximum) also contributed strongly, likely reflecting the association between renal dysfunction and systemic vascular burden. Hemodynamic indicators such as mean SBP and diagnosed hypertension further supported the relevance of circulatory stability in postoperative stroke risk.

Temperature and age were also significant. Lower maximum temperature and advanced age were associated with higher predicted risk, potentially indicating impaired thermoregulation and diminished physiologic reserve. Respiratory rate (both maximum and mean) contributed moderately, underscoring respiratory instability as a marker of systemic stress.

Additional contributors—including BUN, prothrombin time, and heart rate—demonstrated consistent but more modest influence. Elevated BUN and heart rate were generally associated with higher predicted risk, possibly reflecting volume dysregulation or sympathetic activation. Reduced prothrombin time was likewise linked to elevated risk, suggesting that coagulation dynamics may play a secondary role in cerebrovascular vulnerability.

Importantly, unlike prior studies, ICU length of stay was intentionally excluded to avoid information leakage. Nevertheless, the model retained strong predictive performance, suggesting that physiologic and comorbidity features alone sufficiently captured postoperative stroke risk.

Clinically, SHAP values enable patient-specific interpretability by highlighting the key features driving individual stroke risk predictions. This transparency facilitates targeted interventions—such as intensified cardiovascular management in hypertensive patients or enhanced renal monitoring in those with rising creatinine—thus supporting the integration of machine learning into precision ICU decision-making.

\subsection*{Ablation Analysis of Model Predictive Factors}
\begin{figure}[H]
    \centering
    \includegraphics[width=\textwidth]{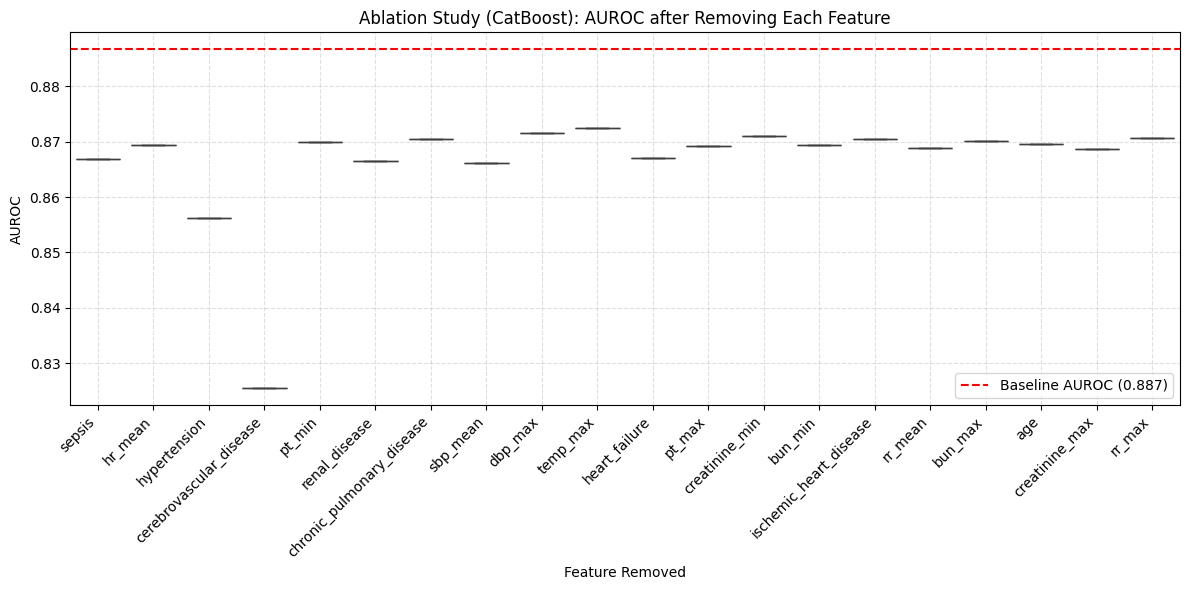}
    \caption{Ablation study of the CatBoost model. Each box represents the distribution of AUROC scores after removing one individual feature. The dashed red line denotes the baseline performance when all features are included.}
    \label{fig:ablation}
\end{figure}

To further assess the stability and interpretability of the CatBoost-based model in predicting postoperative stroke, we conducted a comprehensive feature ablation study. As illustrated in Fig.~\ref{fig:ablation}, each of the top 20 SHAP-ranked features was sequentially removed, and the model was retrained using the selected feature set. Performance was assessed using five-fold stratified cross-validation repeated ten times to ensure statistical consistency. The AUROC of the baseline model trained with all selected features was 0.887, which served as the benchmark for quantifying the marginal contribution of each individual feature.

Figure~\ref{fig:ablation} illustrates the AUROC distribution resulting from the exclusion of individual features. The red dashed line represents the baseline AUROC, providing a reference for quantifying performance decline due to feature removal.

Among the features examined, the exclusion of cerebrovascular disease, minimum creatinine, and mean SBP resulted in the most substantial declines in AUROC, highlighting their central importance in the model's predictive capacity. These variables are clinically meaningful, capturing dimensions of prior cerebrovascular risk, renal function, and hemodynamic stability. Features such as age, maximum BUN, and minimum PT also led to moderate performance drops, suggesting secondary yet notable contributions. Interestingly, variables like sepsis and chronic pulmonary disease—while less prominent in univariate SHAP analysis—nonetheless impacted model performance in the multivariate setting, reflecting complex inter-feature dependencies captured by the CatBoost architecture.

Taken together with the SHAP findings, the consistent decline in AUROC across most ablated features suggests that the model derives predictive strength from an integrated set of physiological, laboratory, and comorbidity-based indicators rather than relying heavily on a single dominant variable. This distributed reliance enhances the model’s resilience to missing or incomplete inputs and supports its generalizability across diverse patient subgroups in real-world ICU settings.

Clinically, the prominence of features such as cerebrovascular disease, creatinine levels, and SBP is consistent with established stroke risk factors, lending physiological credibility to the model’s predictions. By incorporating interpretable and clinically relevant variables, the model demonstrates strong potential for integration into real-time clinical decision-support systems aimed at early risk stratification and proactive intervention.

In summary, the ablation analysis confirms the stability of the predictive framework and underscores the complementary contribution of integrating multiple clinically meaningful features. This multifactorial structure enhances the statistical validity, interpretability, and potential clinical utility of the model in supporting postoperative stroke risk assessment among elderly SICU patients.

\section*{Discussion}
\subsection*{Summary and Review of Existing Models}
This work introduces a reliable and transparent machine learning framework designed to predict postoperative stroke risk in elderly SICU patients, based on detailed clinical data sourced from the MIMIC-III and MIMIC-IV databases. Through a comprehensive preprocessing and modeling pipeline—including missing data imputation, categorical encoding, class imbalance correction using ADASYN, and systematic model comparison—CatBoost was identified as the top-performing algorithm. It achieved an AUROC of 0.8868 (95\% CI: 0.8802–0.8937), outperforming other gradient boosting frameworks such as XGBoost and LightGBM.

The final model relied on a concise and clinically meaningful feature set encompassing key physiologic and comorbidity-related indicators, including cerebrovascular disease, creatinine levels, SBP, age, and temperature. SHAP-based analysis confirmed the central predictive contributions of cerebrovascular and renal variables, reinforcing the biological plausibility and face validity of the model. Stable performance across various metrics and validation folds demonstrates the model’s strong ability to distinguish outcomes and suggests good generalizability in clinical ICU practice.

These results hold significant clinical and operational relevance. Utilizing routinely available data from the initial 24 hours of ICU admission, the model facilitates early recognition of patients at elevated risk. This early risk stratification provides a critical window for timely, targeted interventions—such as hemodynamic stabilization, enhanced neurological assessment, or prompt specialist involvement. Second, the interpretability enabled by SHAP values enhances clinician trust and facilitates actionable insights. For instance, when a patient presents with prior cerebrovascular disease and rising creatinine, the model transparently flags elevated stroke risk, guiding decisions around renal and vascular management. Third, at the health system level, the model could support ICU triage and workflow optimization—especially relevant given rising surgical volumes among aging populations. Importantly, the model’s reliance solely on standard EHR-derived variables supports its broad deployability across diverse care environments without additional infrastructure requirements.

Methodologically, this study benefits from several strengths. It utilizes a large-scale, publicly available ICU dataset containing high-fidelity time-stamped physiologic and laboratory measurements, ensuring reproducibility and transparency. The modeling framework integrates advanced preprocessing techniques to address real-world challenges such as missingness and class imbalance. A multi-model benchmarking strategy, followed by SHAP- and RFECV-based feature selection, further improved predictive performance and interpretability. Model robustness was thoroughly assessed using stratified 5-fold cross-validation, with further support from SHAP-based feature attribution and ablation studies. These validation strategies collectively underscore the model's reliability and clinical applicability in predicting postoperative stroke risk.

\subsection*{Comparison with Prior Studies}
Previous studies have explored machine learning--based approaches to postoperative stroke prediction in elderly patients; however, many have been limited by suboptimal feature engineering, moderate predictive performance, or insufficient model interpretability. Our study addresses these limitations through methodological and algorithmic advancements across data preprocessing, model development, and validation.

Charlesworth et al.\ (2003)~\cite{charlesworth2003stroke} developed and validated a predictive model for perioperative stroke in coronary artery bypass grafting (CABG) patients using logistic regression as a formal predictive modeling approach, rather than purely explanatory analysis. Drawing on data from 33{,}062 patients in the Northern New England Cardiovascular Disease Study Group (1992–2001), they identified seven preoperative predictors—age, sex, diabetes, vascular disease, renal dysfunction (creatinine $\geq$ 2 mg/dL), ejection fraction $<$ 40\%, and urgent/emergent status. The resulting model yielded an AUROC of 0.70 (95\% CI: 0.67--0.72) with near-perfect calibration (observed-to-expected correlation = 0.99), representing one of the earliest structured risk prediction frameworks based on routinely available clinical variables.

Zhang et al.\ (2022)~\cite{zhang2022postoperativestroke} developed a machine learning-based model to predict postoperative stroke in elderly ICU patients using data from the MIMIC-III and MIMIC-IV databases. Unlike traditional statistical analyses aimed at explanatory inference, their study implemented seven supervised learning algorithms, explicitly treating prediction as the primary objective. Among the evaluated models, XGBoost achieved an AUROC of 0.78 (95\% CI: 0.75--0.81), representing the best discriminative performance among the evaluated models in that study. Feature importance analysis identified hypertension, cancer, congestive heart failure, chronic pulmonary disease, and peripheral vascular disease as the most salient predictors. 

In summary, our study leveraged a larger and more contemporary cohort from MIMIC-III and MIMIC-IV and implemented a rigorously structured preprocessing pipeline that addressed missingness, class imbalance, and data heterogeneity in a clinically informed manner. Using a two-stage feature selection approach that combined RF–based importance ranking with recursive elimination, we identified a streamlined set of 20 clinically relevant predictors. This refined feature set maintained predictive performance and was subsequently reviewed and validated by clinical experts to ensure interpretability and practical relevance. The final CatBoost model achieved robust performance, with an AUROC of 0.8868 (95\% CI: 0.8802--0.8937) and an accuracy of 0.9112, reflecting strong discriminative power and stable generalizability across cross-validation folds. To enhance transparency and foster clinical confidence, SHAP-based interpretability analysis was conducted at the global level. The results highlighted physiologically coherent risk factors, such as elevated creatinine, low systolic blood pressure, and pre-existing cerebrovascular disease. Collectively, the advances in data quality, preprocessing architecture, model performance, and explainability contribute to a robust and interpretable risk stratification framework, well-positioned for real-time integration into perioperative stroke prediction workflows.

\subsection*{Limitations and Future Work}
Despite the methodological rigor and enhanced performance of our CatBoost-based model, several limitations merit discussion. First, the study was conducted retrospectively using data from the MIMIC-III and MIMIC-IV databases, which may limit generalizability to other clinical settings and populations. Second, feature extraction was restricted to structured variables within the first 24 hours of ICU admission, potentially overlooking temporal dynamics and evolving physiological signals relevant to stroke risk. Lastly, although the model exhibited strong discrimination and interpretability, it has not yet been externally validated on prospective or multi-center cohorts.

Future work will focus on external validation across diverse ICU settings and the incorporation of time-aware modeling techniques, such as recurrent or transformer-based architectures, to better capture clinical trajectories. We further aim to develop clinician-oriented tools that can be seamlessly embedded within EHR systems, delivering real-time, interpretable decision support for perioperative stroke risk assessment.

\section*{Conclusion}
This study develops and rigorously evaluates an interpretable machine learning framework designed to predict postoperative stroke in elderly SICU patients, based on detailed clinical data from the MIMIC-III and MIMIC-IV databases. Addressing real-world clinical constraints, the pipeline featured stringent cohort selection, SVD-based imputation for missing data, and a modular preprocessing design. Out of an initial set of 80 variables, a clinically relevant subset—including cerebrovascular disease, creatinine, and systolic blood pressure—was identified to balance predictive accuracy with interpretability.

CatBoost achieved an AUROC of 0.8868 (95\% CI: 0.8802--0.8937), representing the highest discriminative performance among the eight classifiers evaluated. The model exhibited a favorable balance between specificity and precision, essential for minimizing false positives in high-risk surgical care. Feature ablation confirmed the marginal importance of top-ranked predictors, while SHAP analysis highlighted the relevance of vascular, renal, and hemodynamic factors in stroke risk stratification.

Clinically, this model enables early identification of high-risk patients using only routinely collected ICU data within the first 24 hours of admission. Its transparency supports clinician trust and promotes integration into real-time decision support systems. Future work will focus on external validation across prospective, multi-center datasets and on extending the framework to incorporate longitudinal and multimodal inputs, such as time-series vitals or neuroimaging. 

Together, these findings establish a robust, interpretable foundation for perioperative stroke risk prediction, offering strong potential for real-time deployment in ICU workflows to improve outcomes in aging surgical populations.

\section*{Acknowledgments}
T.L. independently developed the study concept, designed the research approach, conducted experiments, performed data analysis, and drafted the initial manuscript. S.C. and J.F. assisted with implementation of models and revision of the manuscript. E.P., K.A., and G.P. provided critical feedback on study design and interpretation of findings. M.P. oversaw the project, managed coordination of research efforts, and provided strategic guidance throughout. All authors have reviewed and approved the final version of the manuscript.

We also thank the Laboratory for Computational Physiology at the Massachusetts Institute of Technology for their maintenance of the MIMIC-III database.

\nolinenumbers

%
%
%

\bibliography{Reference_list}     

\end{document}